\documentclass[amssymb,11pt]{article}
\usepackage{amssymb}
\newcommand{\be}{\begin{equation}}
\newcommand{\ee}{\end{equation}}
\newcommand{\ber}{\begin{eqnarray}}
\newcommand{\ear}{\end{eqnarray}}

\newcommand{\de}{\delta}

\newcommand{\ep}{\epsilon}
\newcommand{\et}{\eta}
\newcommand{\fr}{\frac}
\newcommand{\Ga}{\Gamma}

\newcommand{\lb}{\label}

\newcommand{\Lg}{{\cal L}}
\newcommand{\n}{\nonumber\\}

\newcommand{\na}{\nabla}

\newcommand{\st}{\stackrel}

\begin{document}
\title{Strings and Unified Field Theory.}
\author{Mark D. Roberts,\\
{Physikalisches Institut},
{Albert-Ludwigs Universit\"at Freiburg},\\
Herman-Herder Str.3,
Freiburg im Breisgau,
Germany,
D-79104\\
            {mark.roberts@physik.uni-freiburg.de}.
}
\maketitle
\begin{abstract}
It is argued that string theory predicts unified field theory rather than general relativity
coupled to matter fields.  In unified field theory all the objects are geometrical,
for strings the Kalb-Ramond matter field is identical to the nonsymmetric part of the
metric except that the fields contribute to different sides of the field equations.
The dilaton is related to the object of non-metricity.
\end{abstract}
{\tableofcontents}
\section{Introduction.}\label{intro}
By a unified field theory is meant a field theory,  as opposed to an extended object theory,
in which all the fields are geometrical
and occur on the geometric or `left hand side' of gravitational field equations.
If unity is achieved in principle in some form of extended object theory,
why is unity in some form of field theory of interest?
There are two main reasons:
{\it firstly} it is aesthetically more pleasing if a unified description
is still apparent in intermediate descriptions of reality,
{\it secondly} unified field theories as opposed to geometry plus matter theories
give different predictions.

Previous unified field theories have involved torsion $S^a_{bc}\equiv\{^a_{[bc]}\}$,
as this usually is related to fermions or spin it is assumed to vanish here.
Another geometrical object used is non-metricity $-Q_{cab}=g_{ab;c}$,
see \cite{mdr40} and references therein,
typically the object of non-metricity is taken to simplify $Q_{abc}=Q_a g_{bc}$
and $Q_a$ associated with the vector potential $A_a$ of electromagnetism:
the reason such theories break down is that
$Q$ has conformal properties which are not shared with $A_a$.
Yet another approach is to assume that the metric is nonsymmetric and
make some identification such as $g_{[ab]}=F_{ab}$,  see
\cite{einstein45,einstein46,einstein48,einstein50,einstein55,goenner,papapetrou,schrodinger},
where $F_{ab}$ is the faraday tensor of electromagnetism:
the reason such theories typically break down is that $g_{[ab]}$ acts like a
potential tensor rather than a faraday tensor,
see in particular the conclusion of \cite{papapetrou},
in more modern language $g_{[ab]}$ corresponds to the potential
of the Kalb-Ramond \cite{KR}field (classically the same as $2$-form electrodynamics).
Nonsymmetric metrics have been studied recently by
\cite{giannopoulos,hammond,moffat,NS}.
Theories which involve both a nonsymmetric metric
and a scalar dilaton field include \cite{SS}.
There are at least four problems with nonsymmetric metrics:
{\it firstly} the large number of combinations eq.(2)\cite{JP} of ways of constructing
a unified metric,  hopefully the number of possibilities will be reduced by string theory,
{\it secondly} the Weyl \cite{weyl} - Pauli \cite{pauli} objection to Einstein's attempts
which states that because $g_{ab}$ is a reducible representation of diffeomorphisms,
there is no real meaning in saying that a theory is expressed `soley in terms of $g_{ab}$',
{\it thirdly} the Damour - Deser - McCarthy \cite{DDM} problem of the implications of
the gauge invariance of the Kalb-Ramond field for the metric,
this can be overcome by adding a mass term,
and {\it fourthly} the linearization instability of Clayton \cite{clayton},
perhaps this does not occur when the dilaton field is also present.

The present work requires both a non-vanishing object of non-metricity
and an nonsymmetric metric.
This is unlike previous nonsymmetric theories where the vanishing of non-metricity
is one of the primary assumptions.
Here only linear gravitation is considered:  what the full theory could be is left open.
There may be many non-linear theories which give the linearized weak field equations used here.
Even in the simplest cases the relationship between a linear gravitational theory
the corresponding non-linear theory is not immediate \cite{padmanabhan}.
Only states of linearized theories can be compared to those of closed strings,
so that string theory cannot make a direct prediction of which non-linear theory to choose.

In section \S\ref{secqcs} the states of the quantum closed string are derived,
the treatment here comes from \cite{zwiebach}.
In section \S\ref{secwfut} is a discussion of
whether weak unified field theory can be described by
the simplest weak field equations
involving only the d'Alembertian acting on the nonsymmetric linear perturbation $h_{ab}$
and the object of non-metricity $Q$,
when this is the case the theory
has the same one particle states as those of the closed string.
Section \S\ref{seccomp} shows how previous nonsymmetric unified field theories
do not include non-metricity and how Papapetrou's solution \cite{papapetrou}
illustrates how unified field theories
and geometry plus matter theories make different predictions.
Section \S\ref{secconc} is the conclusion.
\section{The quantum closed string.}
\label{secqcs}
The quantum closed string, Ch.13 \cite{zwiebach}, is described by
\be
\sum_{I,J}R_{IJ}a^{I\dagger}_1a^{j\dagger}_1|p^+,\st{\rightarrow}{p}_T>,
\label{qcs}
\ee
where $R_{IJ}$ is an arbitrary square matrix of size $(D-2)$
and $D$ is the dimension of the spacetime.
The sum involving $I$ and $J$ is over spatial indices because in the light cone gauge
components with advanced and retarded null indices can be gauged away.
Roughly speaking an open string has states of the form $a^\dagger|p>$
and linear combinations of these give the general state;
however for the closed string,  apart from an overall momentum,
there are two sets of momenta corresponding to travelling
around the string in either direction,  so that the general state has a two index matrix $R$
which is transvected with two sets of creation operators
corresponding to the two types of momenta.
The arbitrary matrix $R$ can be decomposed into its symmetric $S$,
nonsymmetric $A$ and trace (also called spur) $Sp$ parts.
Applying such a decomposition to (\ref{qcs}) one has states governed by three terms
which are identical to the states of vacuum or stress free linearized gravity,
the Kalb-Ramond field and the dilaton respectively.
\section{Weak field unified theory.}
\label{secwfut}
Assume that there is given an nonsymmetric metric
\be
g_{ab}=g_{(ab)}+g_{[ab]},~~~
g_{(ab)}=\fr{1}{2}(g_{ab}+g_{ba}),~~~
g_{[ab]}=\fr{1}{2}(g_{ab}-g_{ba}).
\ee
Furthermore assume that the determinant of the metric is non-zero.
In odd dimensions the determinant of the nonsymmetric part vanishes.
The inverse of the metric obeys
\be
g_{ab}g^{cb}=\de^c_a,~~~
g_{ab}g^{bc}=\de^c_a+2g_{ab}g^{[bc]}.
\ee
Define the 'Christoffel' connection
\be
\left\{^a_{bc}\right\}\equiv\fr{1}{2}g^{ad}\{g_{bd,c}+g_{cd,b}-g_{(bc),a}\},
\lb{chcon}
\ee
without the symmetrization on the last metric term there is non-vanishing torsion.
Define the contorsion
\be
K^a_{.bc}\equiv g^{ad}\left(-2S_{\{bcd\}}+Q_{\{bcd\}}\right),
\label{contorsion}
\ee
where the Schouten \cite{schouten} bracket is defined by
\be
\{bcd\}\equiv bdc+cdb-bcd.
\label{schoubracket}
\ee
The full connection is the sum of the Christoffel connection and the contorsion
\be
\Ga^a_{bc}\equiv\{^a_{bc}\}+K^a_{bc}.
\label{fullconnection}
\ee
The definitions of the contorsion (\ref{contorsion})
and the full connection (\ref{fullconnection})
sometimes differ by constant factors in various texts.
The torsion and non-metricity are given by
\ber
S^a_{bc}&\equiv&\{^a_{[bc]}\}=\fr{1}{2}k_1g^{ad}g_{[cb],a},\n
-Q_{cab}&\equiv& g_{ab;c}=g_{ab,c}-\Ga^e_{ac}g_{eb}-\Ga^e_{bc}g_{ae}\n
&&=g_{[ab],c}+k_2\left(g_{[bc],a}+g_{[ac],b}\right)-2k_3K_{(ab)c},
\lb{nm}
\ear
the torsion is defined in terms of the Christoffel connection rather than the full connection
so as to avoid it being defined in terms of itself,
when (\ref{chcon}) is defined with symmetrization of the last metric term the torsion vanishes,
i.e. $k_1=0$;  also
$k_2=1$ when the is no symmetrization in the last term of (\ref{chcon})
and $k_2=1/2$ when there is symmetrization.
$k_3=0$ if the connection used in the definition of $Q$ is the Christoffel connection
and $k_3=1$ if the connection used is the full connection.
The Christoffel-Riemann tensor is
\be
\st{\{\}}{R^a}_{.bcd}
=\{^a_{db}\}_{,c}-\{^a_{cb}\}_{,d}+\{^a_{cf}\}\{^f_{db}\}-\{^a_{df}\}\{^f_{cb}\}.
\ee
The symmetries of this tensor are $R_{ab(cd)}=R_{a[bcd]}=0$ in the torsion free case;
the Bianchi identity is complicated but as neither requiring parts of it to vanish
separately, or coupling it to matter are used here it can be ignored.
For any connection which is a sum of the
Christoffel connection and a tensor connection $K^{a}_{.bc}$,  the Riemann tensor is
\be
R^{a}_{.bcd}=\stackrel{\{\}}{R^a}_{.bcd}+\stackrel{K}{R^a}_{.bcd},~~~
\stackrel{K}{R^a}_{.bcd}=
2K^{a}_{.[d|b|;c]}+2K^{a}_{.eb}S^{e}_{dc}+2K^{a}_{.[c|e|}K^{e.b}_{.d]}.
\label{eqra}
\ee
Subject to the weak field approximation
\be
g_{ab}=\et_{ab}+h_{ab},~~~
h_{ab}=h_{(ab)}+h_{[ab]},
\ee
where $h_{ab}$ is the linear perturbation.
and assuming that partial differentiation can be interchanged i.e.$X_{,bc}=X_{,cb}$
the linearized Christoffel-Riemann tensor takes the form
\be
2\st{\{\}}{R^a}_{.bcd}=h^{~a}_{d.,bc}-h^{~~~~,a}_{(bd)~.c}-h^{~a}_{c.,bd}+h^{~~~~,a}_{(bc)~.d}.
\label{linrie}
\ee
There are two different ways to contract to give the Ricci tensor
\be
2\st{\{\}}{R}_{ab}=h_{(ac).b}^{~~~~c}+h_{(bc).a}^{~~~~c}-\Box h_{(ab)}-h_{,ab}
+k_4h_{[bc].a}^{~~~~c}+k_5h_{[ca].b}^{~~~~c}-k_6\Box h_{[ab]},
\label{linricci}
\ee
Contracting (\ref{linrie}) over $a=c$ gives $k_4=1,~k_5=k_6=0$,
contracting over $b=d$ gives $k_4=0,~k_5=k_6=1$,
for the first of these the contribution of $h_{[ab]}$
vanishes in the nonsymmetric harmonic gauge (\ref{ahg}),
so the second choice is taken.
Applying the standard harmonic gauge
\be
h^{~~~~,b}_{(ab)}=\fr{1}{2}h_{,a},
\label{shg}
\ee
to the first and second terms and the nonsymmetric harmonic gauge
\be
h^{~~~~,b}_{[ab]}=0,
\label{ahg}
\ee
to the sixth term
and again assuming that partial differentiation can be interchanged
\be
2\st{\{\}}{R}_{bd}=-\Box h_{bd},
\ee
the same result as for general relativity except that
the nonsymmetric part of the linear perturbation contributes.
For the semi-metric
\be
Q_{abc}=Q_a g_{bc},
\label{semimetric}
\ee
the contorsion part of the Riemann tensor is
\be
\stackrel{K}{R^a}_{.bcd}=Q_{b;c}\de^a_d-Q_{b;d}\de^a_c,
\label{riecontorsion}
\ee
Contracting to form the full linearized Ricci tensor gives
\be
2\st{\Ga}{R}_{ab}=2\st{\{\}}{R}_{ab}+\stackrel{K}{R}_{ab},~~~
\stackrel{K}{R}_{ab}=k_7Q_{a;b}+k_8Q_{b;a}+k_9g_{ab}Q^e_{.;e},
\label{qbd}
\ee
where cross terms $g_{[ab]}Q_c$ are taken to vanish.
The values of $k_7,~k_8$ and $k_9$ depend on which indices are contracted over
and the way in which contorsion and semi-metricity are defined.
For example,  for the contorsion (\ref{contorsion}) and the semi-metricity (\ref{semimetric}),
contracting over $a=c$ gives $k_7=1-D,~,k_8=k_9=0$,
contracting over $b=d$ gives $k_7=1,~k_9=-1$.
The important point is that there are linear terms in non-metricity
which contribute to the linearized Ricci tensor.
For nonmetric theories applying the definition of the covariant derivative to the equation
\be
(\Box^2-M)g_{ab}=0
\label{bgab}
\ee
gives \cite{mdr40}
\be
M(x)+\stackrel{\{\}}{\na}_{a}Q_{.}^{a}+(\ep+D/2-2)Q_{a}Q_{.}^{a}=0,
\label{nnm}
\ee
where $\ep$ depends on the type of $\Box$ operator assumed.
Linearizing,  so that terms ${\cal{O}}(Q^2)$ are discarded,
and assuming $M=0$ and that $Q_a$ is a gradient vector,  gives $\Box Q=0$.
To summarize this section,
the simplest linearized gravitational field equations involving both an nonsymmetric metric
and the object of non-metricity are
\be
\Box h_{ab}=0,~~~
\Box Q=0,
\label{lingrav}
\ee
the first equation coming from the vacuum linearized Ricci equations (\ref{linricci})
and the second term coming from the non-metric equation (\ref{nnm}).
For these equations to be a correct linearization
of a non-linear theory two things must happen.
The {\it first} is that terms of the form (\ref{qbd}) $\st{K}{R}_{ab}$
must not contribute to linearized Christoffel Ricci tensor,
this could be achieved if the field equations $\st{\{\}}{R}_{ab}=0$
are chosen in preference to $\st{\Ga}{R}_{ab}=0$,
or if the terms $\st{K}{R}_{ab}$ vanish by themselves.
The {\it second} is that (\ref{bgab}) is not usually derivable from a given lagrangian,
but assuming (\ref{bgab}) is the easiest way to get to $\Box Q=0$.
Whether there are non-linear theories with these properties is hard to tell,
as no theories with both nonsymmetric metric and object of non-metricity have been studied.
The quantization of the object of non-metricity $Q$
is the same as for a scalar field \cite{zwiebach}Ch.10.4.
The symmetric part of (\ref{linricci}) fourier transformed to momentum space using
the substitution $\na\rightarrow(i/\hbar)p$ gives \cite{zwiebach}eq.10.89,
so that the quantization of the symmetric part of the metric
is the same as for \cite{zwiebach}Ch.10.6.
The equation $\Box h_{[ab]}=0$ is the same as the Euler equation for the Kalb-Ramond field
in the Coulomb gauge.   To see this,  start with the Kalb-Ramond lagrangian
\be
\Lg=-\fr{1}{6}H_{abc}H^{abc},~~~
H_{abc}\equiv B_{bc;a}+B_{ca;b}+B_{ab;c},~~~
B_{ab}=-B_{ba},
\ee
varying with respect to $B$ gives the Euler equation
\be
H^{abc}_{~~~;a}=\Box B^{bc}+B^{ca;b}_{~~~~a}+B^{ab;c}_{~~~~a}=0
\label{ekr}
\ee
The Coulomb gauge
\be
B^{ab}_{~~;b}=0
\label{ckr}
\ee
is the same as the nonsymmetric harmonic gauge (\ref{ahg}),
applying the Coulomb gauge (\ref{ckr}) to the Euler equation (\ref{ekr}) gives
\be
\Box B^{bc}=0,
\ee
which is the same as the nonsymmetric part of the metric in (\ref{lingrav}),
and is quantized in the same way as \cite{zwiebach}prob.14.6.f.
\section{Comparison with other unified field theories.}
\label{seccomp}
The question arises as to whether string theory predicts Einstein's unified theory.
Nonsymmetric theories usually assume vanishing torsion and non-metricity,
for example,  equations (1) and (2) of Einstein and Strauss \cite{einstein46}
are just these requirements.
This implies that string theory can only predict Einstein's unified theory if the
dilaton is not present,  in which case the theory is not a fully unified field theory.
The field equation of Einstein's unified theory are
\be
R_{(ab)}=0,~~~
R_{[ab];c}=0.
\ee
these field equations are different from the Kalb-Ramond field coupled to general relativity
as then there is no stress on the right hand side of the first of these field equations
and the second is not explicitly an equation of $2$-form electrodynamics.
Considering just the pattern of the closed string states
these equally well predict general relativity plus Kalb-Ramond field plus dilaton
or Einstein's unified theory plus dilaton.
Einstein's intension was to unify gravitation with electromagnetism and when restricted
to a brane there are relations between the Kalb-Ramond field
and electromagnetism p.319 \cite{zwiebach} so that from that perspective Einstein's
objective was achieved;  however from a modern perspective without such a geometric setup
Einstein's unified field theory just unifies gravitation with the Kalb-Ramond field.

Papapetrou \cite{papapetrou} first solution is
\ber
ds^2=-\left(1+\fr{q^4}{r^4}\right)\left[1-\fr{2m}{r}-\fr{\lambda}{3}r^2\right]dt^2
+\fr{q^2}{r^2}dtdr-\fr{q^2}{r^2}drdt\n
+\left[1-\fr{2m}{r}-\fr{\lambda}{3}r^2\right]^{-1}dr^2
+r^2(d\theta^2+\sin(\theta)^2d\phi^2).
\label{pap1}
\ear
This is not what would be expected for a solution of $p$-form electrodynamics coupled
to the field equations of general relativity,
apart from the nonsymmetric terms $g_{tr}=-g_{rt}$ one would anticipate
\be
-g_{tt}=\fr{1}{g_{rr}}=
\left[1-\fr{2m}{r}-\fr{\lambda}{3}r^2+\left(\fr{q}{r}\right)^{2p}\right],
\ee
and this illustrates that unified field theories have different solutions
to geometry plus matter theories.
\section{Conclusion.}
\label{secconc}
The only difference between the nonsymmetric part of the metric and the Kalb-Ramond field
is that the nonsymmetric part of the metric occurs on the geometrical side of the field equation
and the Kalb-Ramond field occurs on the matter side.
This result should be robust to most specific unified field theories,
however as illustrated by Papapetrou's first solution this difference is enough for the
predictions of the theories to be different.
The relation between the object of non-metricity and the dilaton is more theory specific,
in particular it requires that the contorsion part of the Ricci tensor does not contribute
to the linearized field equations otherwise $\Box h_{ab}$ couples to it.
So why try to replace the dilaton with the object of non-metricity?
There are two reasons:  the {\it first} is that if a complete unified field theory
is sought then something must correspond to the dilaton,  the choice of torsion is
much worse than non-metricity as torsion is usually related to spin,
the {\it second} is that because of the conformal properties
of the object of non-metricity there will probably be different cosmological predictions.
Another problem is that in the standard picture the various objects all have spin,
so what is the spin of a unified nonsymmetric $h_{ab}$,
the answer is that it can be decomposed into $h_{(ab)}$ and $h_{[ab]}$
and these have spins as before.
In the model here the nonsymmetric metric and object of non-metricity are related (\ref{nm})
and this could occur in general.
\section{Acknowledgements}
I would like to thank
Abraham Giannopoulos,
Hubert Goenner,
David Matravers,
John W. Moffat,
John Stachel \&
Gary B. Tupper
for communications on literature on nonsymmetric metrics,
up until then my literature search was swamped by the role of Einstein in kitsch culture.

\end{document}